\documentclass[amsmath]{aastex702}

\usepackage{url}

\usepackage{graphicx}
\usepackage{color}
\usepackage{siunitx}
\usepackage{hyperref}
\usepackage{longtable}
\usepackage{overpic}

\hypersetup{
	colorlinks,
	citecolor={blue},
	urlcolor={red},
	linkcolor={blue},
}

\usepackage{float}
\usepackage{amsmath}

\begin{document}
	
	\title{Hard X-Ray Quasi-Periodic Pulsations in X-Class Solar Flares Observed by Aditya-L1/HEL1OS}
	
	\author[orcid=0009-0009-3048-0109,sname='Kanaujiya']{Sachin Rajkapoor Kanaujiya}
	
	\affiliation{National Institute of Technology Calicut, Kozhikode-673601, Kerala, India}
	\email[show]{sachin\_p220261ph@nitc.ac.in}

	\author[orcid=0000-0002-6353-9252,sname='Maurya']{Ram Ajor Maurya}
	\affiliation{National Institute of Technology Calicut, Kozhikode-673601, Kerala, India}
	\email[show]{ramajor@nitc.ac.in} 

    \author[orcid=0009-0008-2934-1227,sname='Gopinathan']{Swetha Gopinathan}
	\affiliation{National Institute of Technology Calicut, Kozhikode-673601, Kerala, India}
	\email[show]{swetha\_p250230ph@nitc.ac.in} 
    
	\author[orcid=0000-0001-6423-8286,sname='Nakariakov']{Valery M.	Nakariakov}
    \affiliation{Centre for Fusion, Space and Astrophysics, Department of Physics, University of Warwick, Coventry CV4 7AL, UK}
    \affiliation{G-LAMP NEXUS Institute/School of Space Research, Kyung Hee University,  Yongin, 17104, Republic of Korea}
	\affiliation{Centro de Investigacion en Astronom\'ia, Universidad Bernardo O'Higgins, Avenida Viel 1497, Santiago, Chile}
	\email[show]{V.Nakariakov@warwick.ac.uk}
	
	\shorttitle{QPPs in X-Class Flares}
	\shortauthors{Kanaujiya, et al. 2026}
	
	\begin{abstract}
We present the catalogue of quasi-periodic pulsations (QPPs) detected in hard X-ray (HXR) observations of X-class solar flares obtained with the HEL1OS instrument onboard Aditya-L1 during July 2024--March 2026. The catalogue comprises 34 flares spanning GOES classes X1.1--X7.1. QPPs are detected and studied using a uniform procedure based on Ensemble Empirical Mode Decomposition (EEMD) and wavelet analyses. Statistically significant QPPs are identified in 25 events ($\sim$74\%), demonstrating that QPPs are a common property of X-class flares. The detected periods range from approximately 1 to 3~min, with a pronounced concentration between 1.3 and 1.7~min, suggesting the existence of a preferential timescale that may be associated with kink or slow magnetohydrodynamic oscillations in the flaring region. The dominant periods remain nearly independent of photon energy across the 8--20~keV range, while the modulation depth increases systematically with photon energy following a power-law relation with an exponent of about 1.5. QPPs detected in different HXR energy channels exhibit strong phase coherence and positive cross-correlation coefficients, indicating modulation by a common physical mechanism. Several intense flares display HXR modulation depths of up to $\sim$25\%, providing evidence for nonlinear effects. The observed statistical properties support models in which QPPs are generated by repetitive magnetic reconnection, either spontaneous or periodically modulated by magnetohydrodynamic oscillations. This study establishes HEL1OS as a powerful instrument for quantitative studies of flare QPPs during Solar Cycle~25.

	\end{abstract}
	
	\keywords{Sun: flares; Sun: X-rays; Sun: oscillations }
	
	\section{Introduction}
	\label{sec-intor} 
	
	Quasi-periodic pulsations (QPPs) are an oscillatory phenomenon that has been commonly detected in light curves of energy release events such as solar and stellar flares. QPP are observed to appear within a wide range of timescales, from a fraction of a second to several tens of minutes, and occur in every wavelength range, from radio through optical, EUV, soft and hard X-rays, and even gamma-rays \citep[e.g.,][]{Parks1969, 1987ApJ...321.1031T, Asai2001,Nakariakov2003a, Foullon2005,Ofman2006,Li2008,Srivastava2008,Nakariakov2009,Nakariakov2010,Yuan2011,Yang2016,Hayes2016,Hayes2019,Kupriyanova2020,2026ApJ..1003...78K}.  QPPs often appear simultaneously in thermal emission channels such as EUV and soft X-rays, as well as in non-thermal emissions such as hard X-rays and radio, indicating that they are intrinsic to the physical processes such as flare energy release and particle acceleration.

	QPPs have been reported in the literature over several decades. Tens-of-seconds periodicities in microwave bursts of solar flares are reported as early as the 1960s and 1970s \citep[e.g.,][]{Parks1969,Aschwanden1987}. The Yohkoh mission \citep[e.g.,][]{1991yohk.book.....S} had been a breakthrough, where repetitive pulsations with 10--100~s timescales were detected in major flares\citep{2014CEAB...38..111T, Inglis2015, 2015SoPh..290..115S, 2019AdSpR..64.1100S, 2022ApJ...936...87F}. RHESSI  additionally facilitated systematic detection of hard X-ray multi-period QPPs in X-class events. For example, \citet{Inglis2009} found simultaneous 12~s and 28~s QPPs in an X3.9 flare on 3 November 2003, and attributed them either to different magnetohydrodynamic (MHD) modes of coronal plasma structures in the flare site, or oscillatory reconnection.  In radio, the Nobeyama Radioheliograph observations led to the detection of a number of QPP events. In particular, 15--30~s QPPs in microwave emission produced by the gyrosynchrotron mechanism, showed strong correlations with hard X-ray pulsations, indicating the connection between QPPs and particle acceleration \citep[e.g.,][]{Nakariakov2003a,Tan2010}.
	
	The Solar Dynamics Observatory (SDO) that was launched on 11 February 2010  further advanced the QPP research, providing high-cadence sequences of high-resolution EUV imaging with the Atmospheric Imaging Assembly \citep[AIA,][]{2012SoPh..275....3P,lemen2012aia}. 
     
    The multi-wavelength study by \citet{Simoes2015} showed coherent QPPs in the EUV, soft, and microwave bands, indicating a common driving factor.	More recent studies by \citet{Hayes2019,Hayes2020} using AIA and Fermi/GBM data revealed continued 50--60~s QPPs in X1.6 and X2.0 flares. 
    \citet{2018ApJ...858L...3K} demonstrated the coexistence of 12--25 s and 4--5 min QPPs in the thermal emission of the X9.3-class solar flare that occurred on 6 September 2017, the most powerful flare of Solar Cycle 24, using observations from GOES/XRS and SDO/EVE. Together, these studies demonstrate that QPPs are ubiquitous and even, perhaps, an intrinsic feature of X-class solar flares. Furthermore, \citet{2016ApJ...830..110C} established similarities between QPPs in the soft X-ray emission of solar flares and those of much more energetic stellar flares using observations from RHESSI and XMM-Newton, respectively.
	
	The physical mechanisms responsible for QPPs are an active area of research, with various advanced viable models proposed \citep[see][for comprehensive reviews]{McLaughlin2018, Kupriyanova2020, Zimovets2021}. The broad variety of oscillation periods, oscillation envelopes and other non-stationary properties suggests a co-existence of different processes leading to the oscillatory modulation of flare emission. In general, QPP mechanisms can be divided into three groups. QPPs may be produced by modulation of parameters of the emitting plasma or kinematics of non-thermal electrons accelerated in the flare site (the \lq\lq injection\rq\rq) by an MHD oscillation, \citep[e.g.,][]{2008PhyU...51.1123Z, 2012ApJ...761..134N, 2022MNRAS.516.2292K}. The acceleration itself can be (quasi)-periodic due to modulation of the reconnection rate by an MHD oscillation in the flaring site or nearby \citep[e.g.,][]{Nakariakov2016}, or emanating from the chromosphere \citep[e.g.,][]{Sych2009}. Furthermore, the production of non-thermal electrons and heating of the plasma can appear periodic because of repetitive reconnection. A combination of these mechanisms is probably responsible for the simultaneous occurrence of QPPs with very different oscillation periods in individual flares \citep[e.g.,][]{Kolotkov2015, 2017MNRAS.471L...6L, 2018ApJ...858L...3K, 2021ApJ...921..179L, 2022SoPh..297....2N, 2025JGRA..13033772L}.

	Despite the large number of observational and theoretical investigations, there are a number of open questions related to the QPP phenomenon. The validation of various theories and establishing intrinsic properties of QPP parameters require one to go beyond specific case studies, i.e., to perform ensemble studies using catalogues of QPP events. Such statistic studies have established the ubiquity of QPP in solar flares \citep[e.g.,][]{2015SoPh..290.3625S, 2019AdSpR..64.1100S, 2020ApJ...895...50H}. A similar approach may help us to reveal the dependencies of QPP parameters on the energy of the modulated emission.
	
	A further limitation in the work performed so far is the heterogeneous approach to background trends and noise. The QPP signals are hosted by strongly non-stationary flare light curves. Consequently, estimates of QPP parameters may become significantly biased. In particular, if they are obtained from whole flare time intervals rather than from oscillatory phases. Wavelet techniques are widely used for QPP detection, but restrict analysis to instantaneously harmonic oscillatory signals. However, QPP are often intrinsically non-harmonic \citep[e.g.,][]{Nakariakov2019}. Furthermore, fewer studies explicitly confine the amplitude, or the modulation depth estimates to wavelet-significant intervals. This is necessary to isolate genuine oscillatory power that is not related to background fluctuations.
	 A possible way to avoid shortcomings of the Fourier-based analytical approach, such as the wavelet transform, is to use an analysis technique that specifically addresses the non-stationary nature of the analysed signals \citep[e.g.,][]{2022SSRv..218....9A}. A suitable technique is the Empirical Mode Decomposition \citep[EMD,][]{Huang1998}, which has been applied to QPP analysis, for example, \citet{Kolotkov2015, 2019ApJS..244...44B, 2026ApJ..1003...78K}. In particular, \citet{2016ApJ...830..110C} used EMD to reveal the universality of the linear scaling of the damping time with the oscillation period in QPPs detected in the decay phase of soft X-ray light curves of solar and stellar flares. 
    	
	The HEL1OS instrument \citep{Nandi.Aditya2025} onboard Aditya-L1 \citep{Parate.Aditya2025} offers an opportunity to further advance our understanding of the QPP phenomenon. HEL1OS provides continuous and high-cadence energy-resolved observations of X-ray light curves of solar flares that often host QPP. 
    The aim of this study is to analyse QPPs in an ensemble of X-class flares observed with HEL1OS. Moving beyond individual case studies, we seek to establish a general characterization of hard X-ray QPPs in X-class solar flares. Specifically, we aim to: (i) systematically quantify the dependence of QPP modulation depth on photon energy; (ii) determine whether the dominant QPP period remains invariant across different energy channels; and (iii) investigate possible statistical relationships among QPP modulation depth, period, and duration.	
	
	The paper is organized as follows. Section~\ref{sec-obs-data} describes the  observations used in this study, then outlines the selection of flare events in subsequent Section\ref{sec-selection-events}. Section~\ref{sec-analyses-methods} presents the analysis procedure in detail, including detrending of the flare signals, wavelet-based identification of statistically significant QPPs, and calculation of modulation depths. The results obtained are discussed in Section~\ref{sec-results-discussions}. Section~\ref{sec-summary} summarises our findings.

	\section{The Observational Data}
	\label{sec-obs-data}
	
	This work utilizes X-ray observations provided by the High Energy L1 Orbiting X-ray Spectrometer (HEL1OS) \citep{Nandi.Aditya2025} onboard the Aditya-L1 mission \citep{Parate.Aditya2025}. HEL1OS  monitors the hard X-ray emission from the Sun, mainly emitted during solar flares.  It provides continuous coverage in the energy range of 8--150~keV, using two complementary semiconductor detector systems. The low-energy regime is covered by a pair of Cadmium Telluride (CdTe) detectors, with the sensitivity over 8-70 keV. The higher-energy regime is recorded by two pixelated Cadmium Zinc Telluride (CZT) modules, which have an operational range of 20--150~keV. Multi-channel observations in the 8--20~keV energy range for most events, enabling a systematic comparison of QPP properties across different energy bands.  
	
	The instrument is spectrally calibrated using onboard $^{241}$Am radioactive sources, which provides the 59.5~keV gamma--ray line as a reference. The achieved spectral resolution is about 1~keV at 14~keV for the CdTe detectors and 7~keV at 60~keV for the CZT units. HEL1OS is capable of recording individual photon events with a temporal precision of 10~ms, although the standard science data products are distributed  at a cadence of 1~s.
	
	To complement QPPs detection in HEL1OS data, we also used data from the X-ray Sensor (XRS) onboard Geostationary Operational Environmental Satellites \citep[GOES,][]{2009SPIE.7438E..02C}. The GOES/XRS instrument measures full-disk integrated solar soft X-ray flux in the 0.5--4~\AA\ and 1--8~\AA\ bands with a cadence of approximately 1~s. These observations were also used to compare the oscillatory characteristics of the thermal soft X-ray emission with those detected in the HEL1OS hard X-ray channels.

	\section{Selection of Flare Events}
	\label{sec-selection-events}
	
	We compiled a comprehensive catalogue of X-class solar flares that occurred during the HEL1OS operational period from July 2024 to March 2026 using the GOES flare archive. A total of 41 X-class flares were identified during the rising and peak phases of Solar Cycle 25, reflecting the enhanced level of high-energy solar activity during this time interval. The catalogue spans flare classes from X1.1 to X7.1 and includes both short- to long-duration events, providing a statistically diverse dataset for investigating QPP characteristics.  	
	We found that 34 of these events were observed by HEL1OS with sufficient data quality for QPP analysis, for details see Table~\ref{tab:events-list-qpp-properties}.

	\section{QPP Detection}
	\label{sec-analyses-methods}
	
	We analysed each flare in four HEL1OS energy bands ($8$--$11$, $11$--$14$, $14$--$17$, and $17$--$20$ keV). For each event, the extracted light curve covered the complete flare duration together with an additional 10--20 min before the flare onset and after the decay phase to ensure reliable background characterization. We then applied the Ensemble Empirical Mode Decomposition (EEMD) technique \citep{Wu2009}. EEMD suppresses the mode-mixing effect that limits the performance of the standard EMD technique by averaging an ensemble of signal realizations with added white noise. The resulting decomposition represents the signal as a set of oscillatory components known as Intrinsic Mode Functions (IMFs). We performed 10,000 ensemble trials with a noise width of 0.2, corresponding to Gaussian white noise with a relative amplitude equal to 20\% of the input signal amplitude added to each ensemble trial before decomposition \citep{pyemd}. The decomposition yields a finite set of IMFs, each representing an oscillation with a distinct characteristic timescale.

	Next, we analysed the oscillatory power of IMFs using the Morlet wavelet transform following \citet{Torrence1998}. We evaluated the statistical significance of the wavelet power against a red-noise background model at the 95\% confidence level and regarded only the power exceeding this threshold as evidence of significant QPPs. Although some events were found to exhibit more than one statistically significant IMF, the final selection relied on the coherence and persistence of the oscillatory signal, the strength of the significant wavelet power, and the level of contamination from short timescale fluctuations. Accordingly, we retained the IMF that exhibits the clearest and most coherent oscillatory behaviour for determining the QPP period, duration, phase difference, and quality factor.
	
	The dominant oscillation period was determined from the global wavelet spectrum of the selected IMF. In contrast, the modulation depth and oscillation duration were derived from the reconstructed detrended light curves instead of the isolated IMF. To define the time interval over which these parameters were evaluated, the 95\% wavelet significance mask of the selected IMF was applied, and only the continuous segment with wavelet power exceeding the confidence threshold was retained. This statistically significant segment was treated as the coherent QPP interval for all subsequent measurements. The duration of the oscillation was taken as the total length of this coherent interval. The quality factor  was calculated as the ratio of the duration to the dominant oscillation period \citep[see, e.g.,][]{NakariakovMelnikov2009, 2017A&A...608A.101P}. The modulation depth that characterises the amplitude of intensity fluctuations normalised by instantaneous background emission \citep[e.g.,][]{Nakariakov2003a, Nakariakov2010, Li2024}, was computed exclusively from the reconstructed detrended light-curve samples within the same statistically significant interval, as
\begin{equation}
	\mathcal{M} (\%) =
	\sum\limits_{t}{\frac{\sqrt{[I(t)-I_{\rm 0}(t)]^2}}
		{I_{\rm 0}(t)}}
	\times 100,
	\label{eq-modulation}
\end{equation}
 where $I_{\rm 0}(t)$ represents the instantaneous flare background emission and $I(t)$ denotes the IMF-filtered intensity that contains the QPP signal. We evaluated both quantities over identical wavelet-significant intervals to ensure a consistent determination of the oscillatory amplitude. This normalisation provides a meaningful measure of the QPP strength and enables a quantitative comparison of oscillation amplitudes across different events.

Using a common time interval for both the modulation depth and quality factor calculations, all derived parameters consistently represent the same coherent QPP episode. This procedure eliminates non-oscillatory phases that would otherwise dilute the measured amplitudes and enables consistent comparisons of QPP properties across different energy bands and flare events. 

	\begin{figure*}[b]
		\centering
		\includegraphics[width=0.49\textwidth,clip,viewport=14 15 326 565]{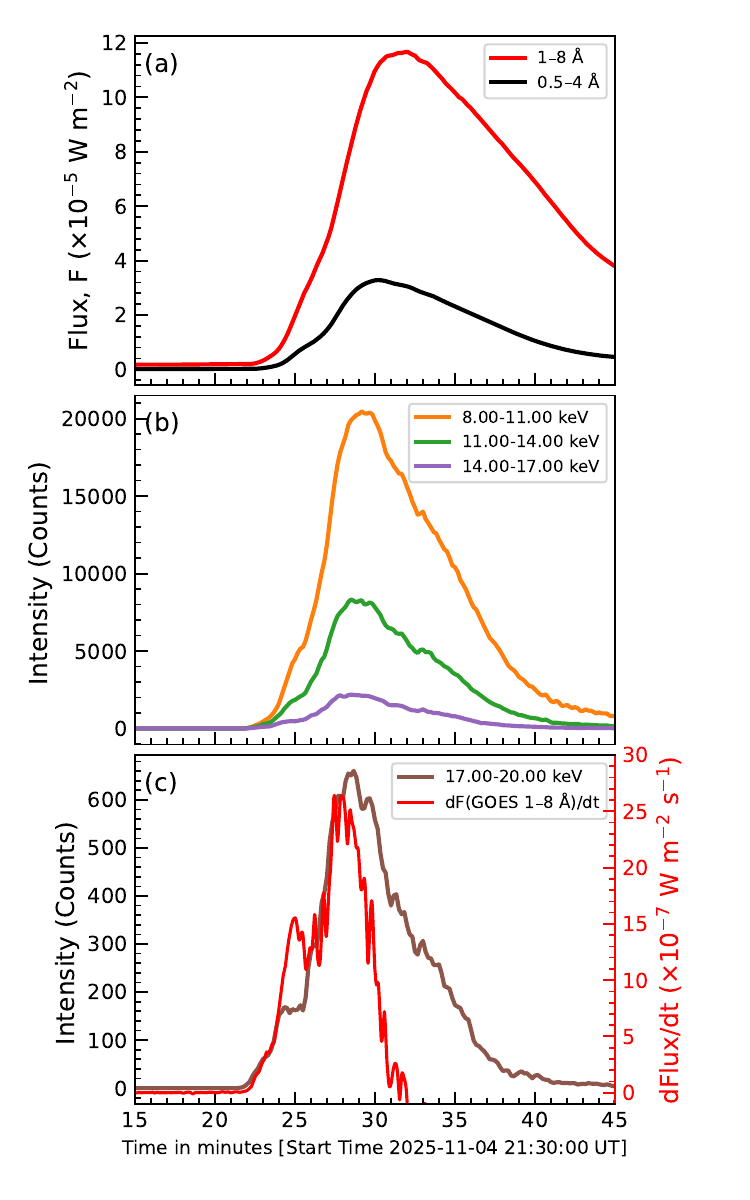}
		\includegraphics[width=0.49\textwidth,clip,viewport=12 15 306 564]{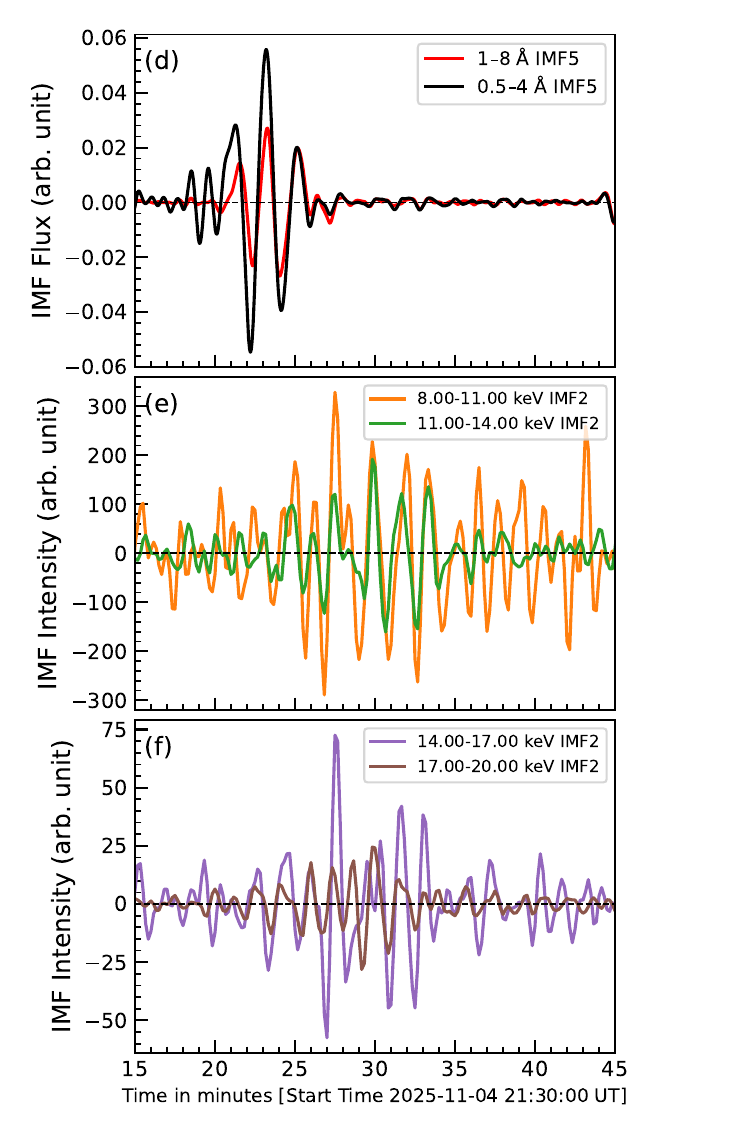}
		\caption{GOES SXR (a) and HEL1OS HXR ((b) -- (c)) flare light curves across different energy channels. Panels (d) -- (f): dominant IMFs corresponding to panels (a) --(c), respectively.  Red curve in panel (c) represents the time derivative of GOES SXR flux in the 1.0--8.0~\AA\ channel.}
		\label{fig:Light_curve_aditya_L1}
	\end{figure*}

	\section{Results and Discussions}
	\label{sec-results-discussions}
	
	The results of our analysis are presented in Figures~\ref{fig:Light_curve_aditya_L1} through~\ref{fig:qpp_statistics} and Tables~\ref{tab:mean-modulation-period} and \ref{tab:events-list-qpp-properties}. These findings are discussed in detail in the following sections.
	
	\subsection{Example: Analysis of the X1.1 Flare on 4 November 2025}
	\label{subsec:x11_flare_analysis}
	
	To demonstrate the standard analytical procedure adopted throughout this study, we first present a detailed multi-wavelength analysis of the X1.1-class solar flare that occurred on 4 November 2025. We systematically applied the same methodology to all remaining flare events in our sample (see Table~\ref{tab:events-list-qpp-properties}). 
	
	\subsubsection{Temporal and Spectral Evolution of X-ray Emission}
	
	Figure~\ref{fig:Light_curve_aditya_L1} summarizes the temporal evolution of the flare emission and the EEMD results used for the QPP analysis. Panel (a) presents the GOES soft X-ray (SXR) light curves in the 1--8~\AA\ and 0.5--4~\AA, channels, while  panels (b) and (c) show the corresponding hard X-ray (HXR) light curves observed by the HEL1OS detectors in the four energy bands spanning 8.00--20.00~keV. The HXR emission exhibits a clear energy-dependent intensity, with the highest count rates in the 8.00--11.00~keV band and progressively lower count at higher energies.
	
	Both SXR channels exhibit a gradual rise, followed by a peak and a subsequent slow decay. In particular, the  1--8~\AA\ channel begins to rise at approximately minute 21.1, reaches its maximum near minute 32.0, and then gradually declines. In contrast, all HXR channels begin and reach thier maximum earlier than the SXR emission. For example, the  17.00-20.00\,keV HXR emission begins at approximately  minute 21.0 and peaks near minute 28.7, approximately 3.3~minute before the SXR maximum.  The temporal correspondence between the HXR emission and the time derivative of the SXR flux (Figure~1(c)) is consistent with the Neupert effect, in which the impulsive energy deposition traced by the HXR emission contributes to the subsequent thermal response of the SXR-emitting plasma. Thus, the observed timing supports a scenario in which accelerated electrons deposit energy in the lower atmosphere and drive the subsequent increase in thermal plasma emission through chromospheric heating and evaporation.
    
    Figure~\ref{fig:Light_curve_aditya_L1}(d--f) display dominant IMFs extracted  from the original light curves  using the EEMD. We selected these IMFs for the subsequent QPP as they  effectively isolate the underlying oscillatory components from the slowly varying, large-scale flare background and high frequency noise, revealing pronounced and coherent quasi-periodic signatures near the flare peak in both the HXR and SXR emissions. \mbox{Panel (e) and (f)} demonstrate a strong energy dependence in the amplitude of these extracted signals. The IMF amplitude scales inversely with the photon energy, exhibiting its maximum peak-to-peak variation in the lowest energy band (8.00--11.00~keV). As the energy threshold increases toward the 17.00--20.00~keV range, the amplitude of the oscillatory signal consistently diminishes. Despite these distinct differences in amplitude, the individual peaks and troughs across all four energy bands remain well synchronized in phase. This precise temporal alignment indicates that the same physical mechanism periodically modulates the accelerating electron population across all energies, rather than acting through independent, energy-segregated drivers.
	
	\begin{figure*}[b]
		\centering
		\includegraphics[width=0.495\textwidth,clip]{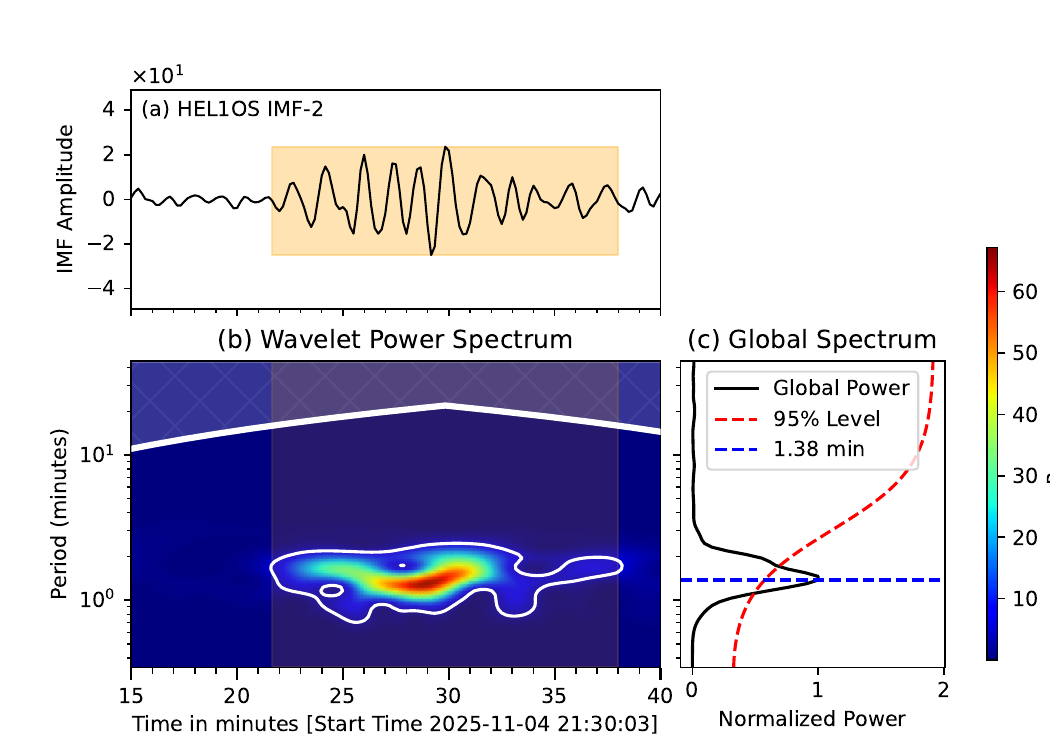}
		\includegraphics[width=0.495\textwidth,clip]{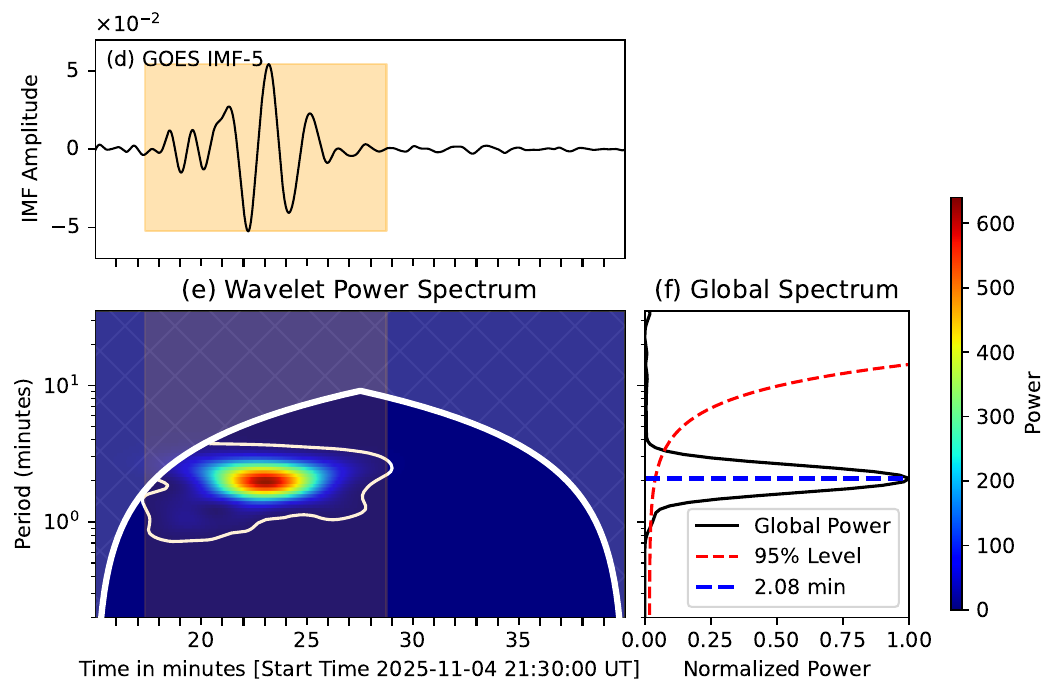}
		\caption{Results of wavelet analysis of the filtered time series extracted from the X1.1-class solar flare on 4 November 2025. Left panels: (a--c) HEL1OS 17--20 keV channel; right panels: (d--f) GOES 0.5--4 \AA\ SXR channel. Panels (a) and (d): selected IMF used for the QPP analysis, with the shaded region indicating the statistically significant interval adopted for the calculation of the modulation depth. Panels (b) and (e): corresponding wavelet power spectra, where the white contour encloses regions exceeding the 95\% confidence level relative to an AR(1) red-noise background, and the hatched region denotes the cone of influence affected by edge effects. Panels (c) and (f): global wavelet spectra with the corresponding 95\% significance levels.}
		\label{fig:wavelet_aditya_L1}
	\end{figure*}
	
	\subsubsection{Wavelet Analysis and Periodicity Detection}
	
	Figure~\ref{fig:wavelet_aditya_L1} presents the Morlet wavelet power spectra along with the global wavelet spectra for the isolated oscillatory components of the flare in HXR (left) and SXR (right) wavelengths. This analysis rigorously establishes periodicities of the QPPs identified in both the HXR and SXR domains, and their statistical significance. Panels (a) and (d) show specific IMF components (HEL1OS IMF-2 for the 17--20~keV channel and GOES IMF-5 for the 0.5--4~\AA\ channel), where the shaded regions define the windows selected for modulation-depth analysis. We restricted the estimation of the modulation depth  exclusively to these shaded intervals where the QPP signals reach their maximum significance and coherence. This restriction minimizes contamination from weak background variations and noise-dominated segments, thereby providing a robust, representative estimate of the pulsation amplitude.
	
	Panels (b) and (e) present the corresponding wavelet power spectra which quantify the temporal evolution of the oscillation power over different periods. For the HEL1OS 17--20~keV channel, the spectrum reveals a highly concentrated, intense power core between minutes 22 and 33.5. A solid white contour encloses this core, demonstrating that the oscillation power significantly exceeds the 95\% confidence level relative to a standard autoregressive model of order 1, AR(1), red-noise background. Similarly, the GOES 0.5--4~\AA\ wavelet power spectrum (panel e) exhibits strong, distinct power localized around minute 23 (spanning minutes 18--28). This primary oscillatory power clusters notably earlier in the flare timeline and remains completely unaffected by edge effects. 
	
	Panels (c) and (f) display global wavelet spectra obtained by averaging the wavelet power over time. The dashed curves represent the 95\% significance levels derived from a red-noise background model following \citet{Torrence1998}. The global power peak for HEL1OS (panel c) exhibits a narrow, well-defined profile that sharply exceeds the 95\% significance threshold, yielding a dominant HXR pulsation period of $1.38 \pm 0.39$ min. The global wavelet spectrum for the GOES channel (panel f) identifies a statistically significant primary period of $2.08 \pm 0.48$ min. Within these uncertainties, the HXR and SXR oscillation periods are approximately consistent with each other.

	\subsection{Properties of Detected QPPs}
    
Table~\ref{tab:events-list-qpp-properties} summarizes the QPP characteristics that we derived from HEL1OS observations of X-class solar flares between September 2024 and February 2026. This survey analyzes QPP signatures across 34 X-class flares spanning GOES classes $X1.0$--$X9.0$. We detected coherent QPP signatures in 25 of these flares ($\sim$74\%), whereas nine events (marked as ``No QPP'' in Table~\ref{tab:events-list-qpp-properties}) lacked statistically significant periodic modulations detected by our technique. This detection rate confirms that QPPs constitute at least a common feature of impulsive hard X-ray emission during major flares. 

\begin{figure*}[b]
		\centering
		\begin{overpic}[width=0.485\textwidth,clip,viewport=8 9 420 276]{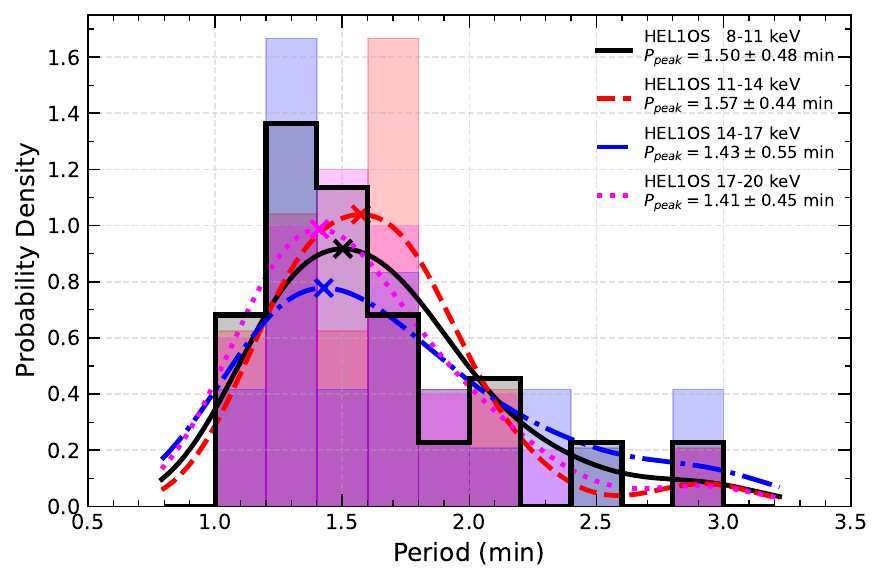}
				\put(15,50){\colorbox{white}{\bfseries (a)}}
		\end{overpic}
		\begin{overpic}[width=0.485\textwidth,clip,viewport=8 9 422 275]{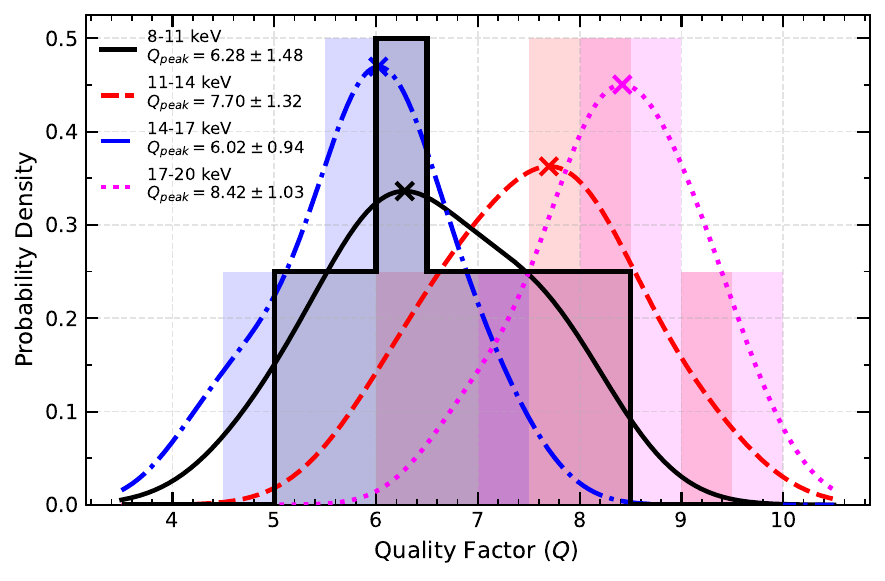}
			\put(85,50){\colorbox{white}{\bfseries (b)}}
		\end{overpic}
			
		\caption{Normalized histograms of the dominant QPP periods (a) and quality factors (b) obtained from the HEL1OS observations of X-class solar flares. The semi-transparent histogram bars correspond to the four analysed hard X-ray energy channels: black (8--11~keV), red (11--14~keV), blue (14--17~keV), and magenta (17--20~keV). The overlap of the bars of different colours reflects the similarity of the corresponding distributions. Solid curves show the kernel density estimates for each energy channel, and crosses indicate the peak of each distribution.}
		
		\label{fig:pdf-periods-qualityfactor}
	\end{figure*}

Figure~\ref{fig:pdf-periods-qualityfactor} presents normalized histograms of these quantities for the four analysed hard X-ray energy channels, together with the corresponding kernel density estimates, which provide a continuous representation of the underlying distributions. The histograms of the dominant periods, shown in panel (a), exhibit very similar behaviour in all four energy channels, with the majority of the measurements concentrated between approximately 1.3 and 1.7~min. The kernel density estimates reproduce this common distribution and show maxima occurring at nearly the same period in each energy band. The strong overlap between the distributions indicates that the characteristic QPP timescale remains essentially unchanged over the investigated hard X-ray energy range and shows little dependence on photon energy.

Normalized histograms of the quality factor are shown in Fig.~\ref{fig:pdf-periods-qualityfactor}(b). Compared with the period distributions, the quality-factor histograms are more scattered, reflecting a wider range of oscillation coherence among individual flare events. Although the distributions overlap substantially for all four energy channels, the kernel density estimates reveal only small differences in their central values, with no evidence for a systematic variation with photon energy. Most of the measured quality factors fall between approximately 5 and 9, indicating that the detected QPPs typically persist over several oscillation cycles. A few flares, such as the X1.8 event on 4 November 2025, display exceptionally large quality factors exceeding 15, suggesting highly coherent periodic energy release. On the other hand, several large flares, including the X9.0 event on 3 October 2024, exhibit QPPs with relatively small quality factors despite their high intensity. The broader spread of the quality-factor distributions is possibly expected, because it is defined as the ratio of the QPP duration to the dominant period, and therefore its uncertainty includes the combined uncertainties associated with both quantities.

Our findings show that the duration of a QPP pattern varies significantly across events, ranging from $\sim$5~minutes in compact flares to over 30~minutes in extended events; longer durations generally accompany higher variations in the modulation depth. The flare intensity displays no clear relationship with the occurrence or properties of QPPs. For instance, strong flares, including the X7.1, X9.0, and X4.0 events, exhibit well-defined QPP patterns, whereas other energetic flares, such as the X5.1 and X8.1 events, show no detectable QPP signatures. Conversely, several relatively weak X1-class flares display highly coherent oscillations with large quality factors. This lack of a monotonic relationship indicates that QPP parameters are determined by a combination of various parameters, such as, possibly, the magnetic configuration and plasma environment, and also the reconnection regime, rather than the flare class alone.

	\begin{figure*}[t]
		\centering

		\begin{overpic}[width=0.30\textwidth,clip,viewport=5 6 348 275]{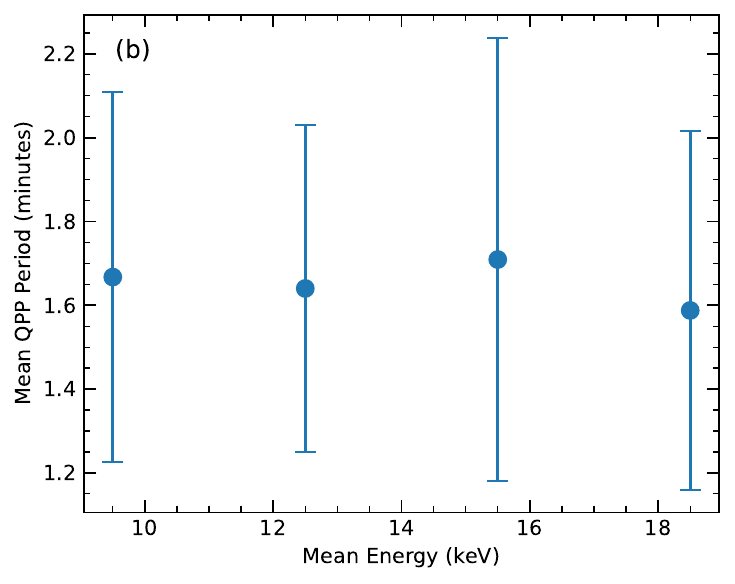}
			\put(15,70){\colorbox{white}{\bfseries (a)}}
		\end{overpic}
		\begin{overpic}[width=0.325\textwidth,clip,viewport=8 10 280 210]{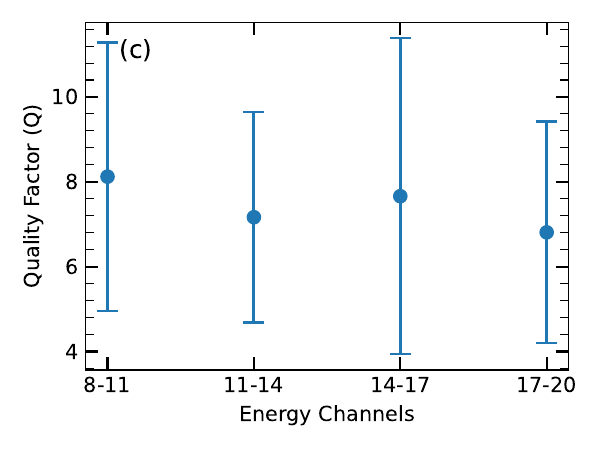}
			\put(17,65){\colorbox{white}{\bfseries (b)}}
		\end{overpic}
		\begin{overpic}[width=0.30\textwidth,clip,viewport=5 5 348 275]{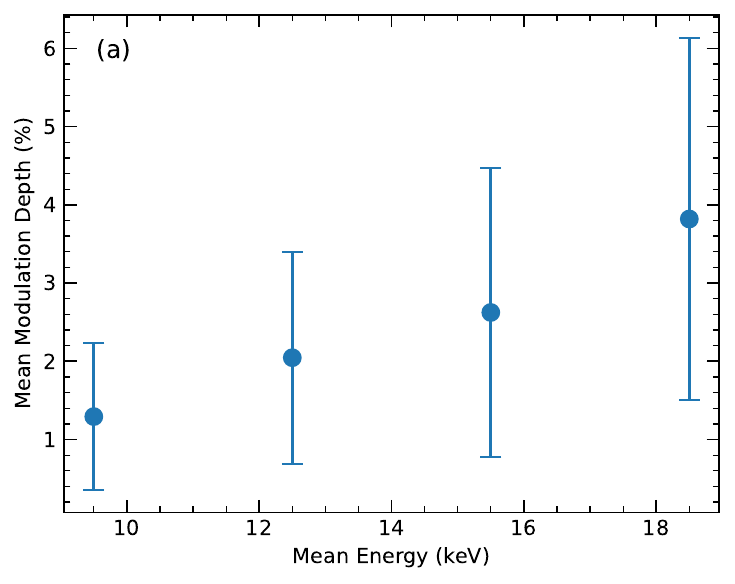}
			\put(12,70){\colorbox{white}{\bfseries (c)}}
		\end{overpic}
		
		\caption{Mean QPP properties derived from the combined HEL1OS flare sample. Panel (a) shows the mean modulation depth as a function of photon energy. Error bars represent one standard deviation within each energy channel. Panel (b) shows the mean QPP period as a function of photon energy. Panel (c) shows the mean quality factors measured in the four HEL1OS energy channels. The error bars indicate the corresponding standard deviations of the measured periods. }
		\label{fig:mean_energy_dependence}
	\end{figure*}
	
	\subsection{Energy Invariance of QPP Periods and Quality Factors}
    
Figure~\ref{fig:mean_energy_dependence}(a) illustrates that the mean QPP period remains remarkably stable across the entire observed energy spectrum. The average periods cluster consistently between 1.6 and 1.7 minutes, yielding specific values of $1.67 \pm 0.44$~minutes in the 8--11~keV band, $1.64 \pm 0.39$~minutes in the 11--14~keV band, $1.71 \pm 0.53$~minutes in the 14--17~keV band, and $1.59 \pm 0.43$~minutes in the 17--20~keV band. The substantial overlap among the standard deviations confirms the absence of systematic energy scaling. This energy-independence suggests a unified, energy-insensitive pacemaker for the oscillations.

Figure~\ref{fig:mean_energy_dependence}(b) presents the quality-factor ($Q$) distributions across the four individual HEL1OS channels. The mean values remain highly stable and comparable across the spectrum, averaging $8.12$ at 8--11~keV, $7.17$ at 11--14~keV, $7.66$ at 14--17~keV, and $6.80$ at 17--20~keV. The broad overlap of these distributions shows that the coherence of the oscillatory signal remains uniform throughout the analyzed energy range.

			\begin{table*}[h]
		\caption{Mean modulation depth $\bar{\mathcal{M}}$ and mean QPP period $\bar{P}$ derived for the HEL1OS hard X-ray energy channels and GOES soft X-ray channels. The quoted uncertainties correspond to one standard deviation of the distributions within each channel.}
		\label{tab:mean-modulation-period}
		
		\centering
		
		\begin{tabular*}{0.90\textwidth}{@{\extracolsep{\fill}} c c c}
			\hline
			\textbf{Energy Channels} &
			\textbf{$\bar{\mathcal{M}} \pm \sigma_{\bar{\mathcal{M}}}$} &
			\textbf{$\bar{P} \pm \sigma_{\bar{P}}$} \\
			&
			\textbf{(\%)} &
			\textbf{(minutes)} \\
			\hline
			
			HEL1OS 8--11 keV   & $1.29 \pm 0.94$ & $1.67 \pm 0.44$ \\
			HEL1OS 11--14 keV  & $2.05 \pm 1.36$ & $1.64 \pm 0.39$ \\
			HEL1OS 14--17 keV  & $2.62 \pm 1.85$ & $1.71 \pm 0.53$ \\
			HEL1OS 17--20 keV  & $3.82 \pm 2.32$ & $1.59 \pm 0.43$ \\
			GOES 1--8 \AA      & $0.36 \pm 0.20$ & $0.94 \pm 0.59$ \\
			GOES 0.5--4 \AA    & $0.48 \pm 0.45$ & $0.94 \pm 0.66$ \\
			
			\hline
		\end{tabular*}
		
	\end{table*}

	\subsection{Energy Dependence of Modulation Depth}
	
Figure~\ref{fig:mean_energy_dependence}(c) shows that the RMS modulation depth increases monotonically with the photon energy across the analyzed HEL1OS channels. The mean modulation depth increases from $1.29 \pm 0.94\%$ in the 8--11~keV band to $2.05 \pm 1.36\%$ in the 11--14~keV band, and reaches $2.62 \pm 1.85\%$ and $3.82 \pm 2.32\%$ in the 14--17~keV and 17--20~keV bands, respectively. Although substantial event-to-event variability exists within each energy channel, the positive trend remains clear. A statistical analysis of the combined dataset yields a Pearson correlation coefficient of $r = 0.48$ ($p = 1.1 \times 10^{-6}$) and a Spearman rank correlation coefficient of $\rho = 0.54$ ($p = 2.3 \times 10^{-8}$), indicating a reliable positive relationship between the modulation depth and photon energy.
	
	\subsection{Cross-Channel Correlation and Multiband Coherence}
	
\begin{figure*}[h]
		\centering
		\includegraphics[width=0.36\textwidth,clip,viewport=8 10 280 208]{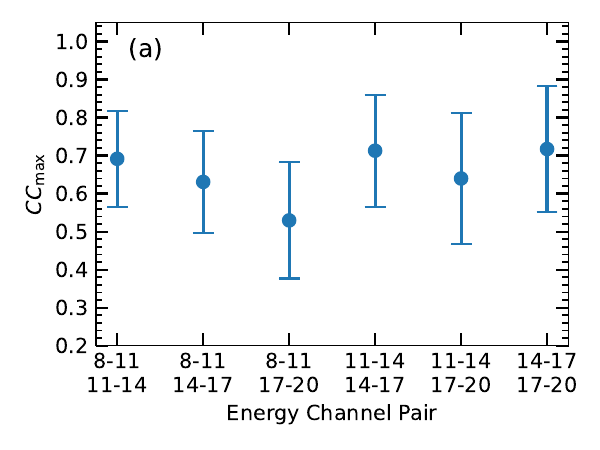}
		\includegraphics[width=0.36\textwidth,clip,viewport=8 10 280 208]{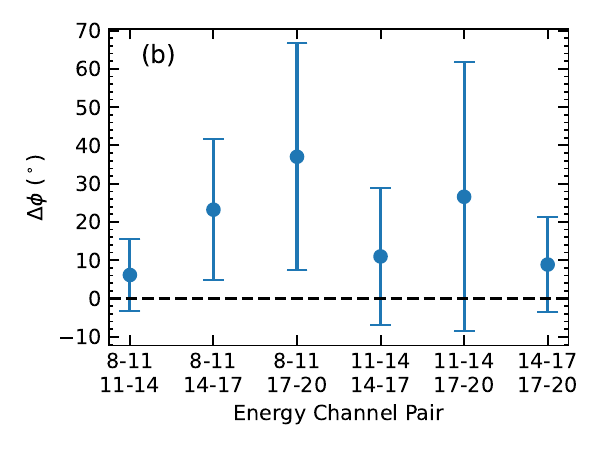}
		\caption{Statistical properties of the detected QPPs. Panel (a) shows the mean maximum cross-correlation coefficients for different pairs of HEL1OS energy channels. Panel (b) presents the corresponding phase differences derived from the oscillatory components. The dashed line indicates zero phase difference.  Error bars represent one standard deviation of the corresponding distributions.}
		\label{fig:qpp_statistics}
	\end{figure*}

	To assess the temporal coherence of the QPPs detected in different energy channels, we examine the maximum cross-correlation coefficients ($CC_{\rm max}$) between QPP pattern for different energy channel pairs, see Figure~\ref{fig:qpp_statistics}(a). The mean $CC_{\rm max}$ values remain consistently high across all six pairs, ranging between approximately 0.53 and 0.72. Adjacent energy bands exhibit the strongest correlation, where the (11--14, 14--17)~keV pair achieves the highest coefficient ($\sim$0.72), followed closely by the (14--17, 17--20)~keV pair ($\sim$0.71) and the (8--11, 11--14)~keV pair ($\sim$0.69). However, cross-correlation systematically degrades as the energetic separation between the compared bands increases, dropping to $0.63$ for the (8--11, 14--17)~keV pair and reaching a minimum of $0.53$ for the widely separated (8--11, 17--20)~keV pair. This high cross-correlation indicates that the same driver modulates the plasma emission across the entire observed X-ray energy spectrum, while energy-dependent transport and emission processes progressively introduce localized decoherence at wider energy gaps.

	\subsection{Temporal Delays and Phase Lag Distributions}
	
	We evaluated the temporal relationships across the energy spectrum using time-lag and phase-difference measurements. Time-lag analyses demonstrate that oscillations remain nearly simultaneous across the observed channels, as most channel pairs exhibit zero time lag within the 10~s temporal resolution of HEL1OS. Delays of 10--20~s occur rarely, and lags exceeding 30~s are exceptionally rare.

	Figure~\ref{fig:qpp_statistics}(b) presents the corresponding phase differences ($\Delta\phi$), which we calculated using the Hilbert transform \citep{Rosenblum1996}. The mean phase differences remain consistently positive and scale with the energetic separation between the paired channels. The phase shift grows from $6.15^{\circ}$ for the adjacent (8--11, 11--14)~keV pair to $23.22^{\circ}$ for the (8--11, 14--17)~keV pair, and peaks at $37.04^{\circ}$ for the most widely separated (8--11, 17--20)~keV pair. Although considerable event-to-event scatter causes individual distributions to overlap the zero-phase line, the ensemble averages demonstrate a systematic energy-dependent time lag where low-energy HXR oscillations precede high-energy oscillations.

	\subsection{Comparison of QPP Modulation in Thermal and Non-thermal Emission}
	\label{sec:goes_comparison}
	
	To compare the behaviour of QPPs in thermal and non-thermal emissions, Table~\ref{tab:mean-modulation-period} evaluates the events that HEL1OS and GOES simultaneously recorded. The GOES soft X-ray channels display relatively small mean modulation depths of $0.36 \pm 0.20\%$ in the 1--8~\AA\ band and $0.48 \pm 0.45\%$ in the 0.5--4~\AA\ band. In comparison, the non-thermal HEL1OS 17--20~keV channel exhibits a substantially larger mean modulation depth of $3.82 \pm 2.32\%$. Despite this marked disparity in relative amplitude, the characteristic periodicities remain comparable across both instruments, yielding a mean period of  1--2~min across both the GOES and HEL1OS observations.

	\section{Summary and Conclusions}
	\label{sec-summary}
	
This study presents a catalogue of QPPs detected in the HEL1OS HXR observations of X-class solar flares during the July 2024--March 2026 interval of Solar Cycle~25. The catalogue includes 34 X-class flares spanning classes X1.1 to X7.1. By applying a uniform analysis procedure based on EEMD and wavelet transforms to all events, we establish a consistent observational framework for investigating the statistical properties of HXR QPPs. This standardized catalogue enables meaningful comparisons between events and provides a rigorous foundation for robust statistical analysis. We detected statistically significant QPPs in 25 flares ($\sim$74\%). This occurrence rate is consistent with the $\sim$80\% reported for X-class flares in Solar Cycle~24 by \citet{Simoes2015}. On the other hand, it is higher than the 46\% occurrence rate reported by \citet{2020ApJ...895...50H}. This discrepancy is most likely due to differences in the QPP detection techniques, as our approach is capable of identifying strongly non-stationary oscillatory patterns. The quality factor which quantifies the duration of a QPP pattern in units of the oscillation period, ranges from approximately 3.1 to 18.9. The majority of events exhibit moderate quality factors of 4--10.
	
We summarise the main conclusions of this study as follows:

\textit{Existence of a preferential period:} The detected QPP periods range from approximately 1 to 3~min, with most values between 1.3 and 1.7~min. The concentration of periods within this interval may indicate a preferential timescale associated with kink or slow magnetoacoustic oscillations of coronal loops in the flaring region. It has been theoretically proposed that the waves can periodically trigger magnetic reconnection or modulate the reconnection rate \citep{2006SoPh..238..313C, 2006A&A...452..343N}. For example, \citet{2016ApJ...822....7K} applied this mechanism to interpret a multi-wavelength QPP event. The detected QPP could also result from a spontaneous repetitive reconnection process \citep[see, e.g.,][]{2000A&A...360..715K, 2025ApJ...980..158K}, while it is not clear what would determine the oscillation period in that scenario \citep{McLaughlin2018}.  Our finding differs from the period distribution reported by \citet{McLaughlin2018}, who compiled a histogram of 278 QPP events published in the literature, showing a peak at 10--60~s. One possible explanation is the different flare populations considered: whereas the compilation of \citet{McLaughlin2018} includes flares spanning all GOES classes, our sample consists exclusively of X-class flares.

\textit{Energy-Invariant Global Timescale:} The dominant pulsation periods cluster around $\sim 1.3$~minutes across all HEL1OS energy bands,  (8--11, 11--14, 14--17, and 17--20 keV). This suggests a common driver that periodically modulates the acceleration or kinematics of non-thermal electrons with those energies.

\textit{Scaling of Amplitude:} Modulation depth of QPP has a positive power-law correlation with the photon energy, having an exponent value of $\alpha\approx 1.4-1.6$. The fitted correlation is highly significant (false alarm probability, $p=0.005$, $R^{2}=0.991$), demonstrating that the periodic mechanism directly modulates the accelerated non-thermal electron population rather than the bulk thermal plasma. 

\textit{High Cross-Channel Coherence:} QPP signals across different HEL1OS energy channels display strong phase synchronisation and consistently positive cross-correlation coefficients. The strongest correlations occur between neighbouring energy bands. There is a small energy-dependent time lag of $37^{\circ}\pm 27^{\circ}$, where low-energy HXR oscillations precede
high-energy oscillations, which requires further investigation. Anyway, these findings indicate that a unified physical mechanism drives QPPs across the entire HXR emission spectrum.

\textit{Enhanced Modulation of Non-thermal Emission:} Comparative analysis with GOES SXR data shows that HXR modulation depths are systematically larger than their thermal SXR counterparts.

\textit{Evidence for Nonlinear Effects:} Several intense flares observed during May--June 2025 exhibit large modulation depths, reaching up to $\sim 25\%$ in the 17--20~keV channel. This finding indicates the role of nonlinear processes in the QPP phenomenon. 

The growth of the modulation depth with photon energy aligns with previous HXR and multi-wavelength QPP investigations that reported enhanced pulsation signatures at higher energies \citep{Li2008, NakariakovMelnikov2009, Kupriyanova2010, Hayes2016, Li2017}. Comprehensive reviews of solar and stellar QPPs similarly emphasise that such energy-dependent amplitudes naturally arise when a repetitive reconnection process results in periodic injection of energetic electrons \citep{Kupriyanova2020, McLaughlin2018}. The energy-independent behavior of the dominant periods matches earlier observational findings of similar QPP periods across a wide range of observational wavelengths \citep{2010ApJ...708L..47N, Inglis2015, Inglis2023}. These findings suggest that QPPs are likely to be produced by repetitive magnetic reconnection, either occurring spontaneously or periodically driven by an MHD oscillation, such as a kink or slow mode. In both scenarios, the principal challenge is to explain the preferential occurrence of oscillations with periods in the 1--3~min range.

In conclusion, this catalogue establishes HEL1OS as a high-fidelity instrument for quantitative QPP diagnostics during Solar Cycle~25. Statistical scalings of various observables create an empirical foundation for advanced theoretical modelling of the QPP phenomenon.

	\section{Acknowledgment}
	
	Aditya-L1 is an observatory class mission which is fully funded and operated by the Indian Space Research Organization (ISRO). The mission was conceived and realised with the help from various ISRO centres. The science payloads and science ready data products are realised by the payload PI institutes in close collaboration with ISRO centres. The PI institutes are: Indian Institute of Astrophysics (IIA); Inter University Centre for Astronomy and Astrophysics (IUCAA), Laboratory for Electro-optics Systems (LEOS/URSC); Physical Research Laboratory (PRL); U R Rao Satellite Centre (URSC); and Space Physics Laboratory (SPL/VSSC). We acknowledge the use of data from the Aditya-L1 mission of the Indian Space Research Organisation (ISRO), archived at the Indian Space Science Data Centre (ISSDC). One of the authors, R.~A.~M., gratefully acknowledges support from the ISRO RESPOND project (No. ISRO/RES/2/437/21-22). V.M.N. is supported by ERC grant 101201424 (ACDCSUN) and the BK21 FOUR program through the National Research Foundation of Korea (NRF) under the Ministry of Education (MoE) (Kyung Hee University, Human Education Team for the Next Generation of Space Exploration), and the Global-Learning \& Academic research institution for Master’s/PhD students, the Postdocs (G-LAMP) Program of the NRF grant funded by the MoE (RS-2025-25442355). 
	
	\appendix
	
	\section{Comprehensive Catalog of QPP Parameters in X-class Flares Detected with HEL1OS }
	
	Table~\ref{tab:events-list-qpp-properties} presents the complete catalog of QPP parameters derived from HEL1OS observations of X-class solar flares during 2024--2026. For each detected QPP event, the table lists the flare class, flare index (FI), energy channel, dominant oscillation period, RMS modulation depth, and quality factor ($Q$). The table also gives the strongest inter-channel coherence properties obtained from the cross-correlation analysis, including the energy-channel pair with the maximum correlation coefficient ($CC_{\rm max}$), the corresponding temporal lag (Lag), and the phase difference ($\Delta\phi$ in degree) between the correlated signals.
	
	These parameters provide a comprehensive characterization of the detected QPPs and their coherence across the HEL1OS energy bands. Most events exhibit small phase differences and negligible temporal lags, indicating that the pulsations remain nearly synchronized across the investigated hard X-ray energy range. The quality factors further show that the detected oscillations are sufficiently coherent to persist over multiple cycles, supporting the robustness of the identified periodicities.

	\small
\setlength{\tabcolsep}{2pt}

\begin{longtable}{|c|c|p{0.9cm}|c|cc|c|c|c|c|p{2.2cm}|c|c|c|}
	\caption{Comprehensive catalog of QPP parameters derived from HEL1OS observations of X-class solar flares in 2024--26.}
	\label{tab:events-list-qpp-properties}\\
	
	\hline
	\multicolumn{4}{|c|}{\textbf{Event}} &
	\multicolumn{2}{c|}{\textbf{GOES/XRS}} &
	\multicolumn{4}{c|}{\textbf{HEL1OS}} &
	\multicolumn{4}{c|}{\textbf{Cross-correlation}}\\
	\hline
	
	S.N. &
	Date &
	Time &
	Flare &
	\multicolumn{2}{c|}{1--8 \AA} &
	Energy &
	P &
	MD &
	$Q$ &
	Channel pair &
	$CC_{\rm max}$ &
	Lag &
	$\Delta\phi$\\
	
	\cline{5-6}
	\cline{11-14}
	
	&&&&
	P (min) & MD &
	(keV) &
	(min) &
	(\%) &
	&
	&
	&
	(s) &
	($^\circ$)\\
	\hline
	\endfirsthead
	
	\multicolumn{14}{c}{\tablename\ \thetable{} -- continued from previous page}\\
	\hline
	
	\multicolumn{4}{|c|}{\textbf{Event}} &
	\multicolumn{2}{c|}{\textbf{GOES/XRS}} &
	\multicolumn{4}{c|}{\textbf{HEL1OS}} &
	\multicolumn{4}{c|}{\textbf{Cross-correlation}}\\
	\hline
	
	S.N.&Date&Time&Flare&
	\multicolumn{2}{c|}{1--8 \AA}&
	Energy&
	Period&
	RMS MD&
	$Q$&
	Energy channels&
	$CC_{\rm max}$&
	Lag&
	$\Delta\phi$\\
	
	\cline{5-6}
	\cline{11-14}
	
	&&&&
	P&MD&
	(keV)&
	(min)&
	(\%)&&
	&&
	(s)&
	(deg)\\
	\hline
	\endhead
	
	\hline
	\multicolumn{14}{r}{Continued on next page}\\
	\hline
	\endfoot
	
	\hline
	\endlastfoot
	
	1 & 12 Sep 24 & 09:31--09:51 & X1.3 &
	0.66 & 0.29 &
	17--20 &
	1.23 &
	3.11 &
	6.66 &
	&
	&
	&
	\\
	\hline
	
	2 & 14 Sep 24 & 15:13--15:47 & X4.5 & 1.56
	& 0.53 & 
	8--11 &
	1.09 &
	0.51 &
	13.11 &
	(8--11,11--14) &
	0.716 &
	0 &
	6.32\\
	
	& & & &
	& &
	11--14 &
	1.16 &
	0.74 &
	8.35 &
	(8--11,14--17) &
	0.618 &
	0 &
	34.37\\
	
	& & & &
	& &
	14--17 &
	2.19 &
	1.06 &
	6.40 &
	(8--11,17--20) &
	0.462 &
	0 &
	65.19\\
	
	& & & &
	& &
	17--20 &
	1.95 &
	3.18 &
	4.11 &
	(11--14,14--17) &
	0.729 &
	0 &
	22.62\\
	
	& & & &
	& &
	&
	&
	&
	&
	(11--14,17--20) &
	0.611 &
	0 &
	16.61\\
	
	& & & &
	& &
	&
	&
	&
	&
	(14--17,17--20) &
	0.759 &
	0 &
	-9.09\\
	\hline
	
	3 & 1 Oct 24 & 21:58--22:29 & X7.1 & 0.58
	& 0.19 &
	8--11 &
	1.30 &
	0.54 &
	8.33 &
	(8--11,11--14) &
	0.606 &
	0 &
	-1.10\\
	
	& & & &
	& &
	11--14 &
	1.30 &
	0.72 &
	6.28 &
	(8--11,14--17) &
	0.601 &
	0 &
	9.93\\
	
	& & & &
	& &
	14--17 &
	1.30 &
	0.89 &
	8.85 &
	(8--11,17--20) &
	0.563 &
	10 &
	32.36\\
	
	& & & &
	& &
	17--20 &
	1.30 &
	0.91 &
	6.28 &
	(11--14,14--17) &
	0.687 &
	0 &
	18.24\\
	
	& & & &
	& &
	&
	&
	&
	&
	(11--14,17--20) &
	0.667 &
	10 &
	23.95\\
	
	& & & &
	& &
	&
	&
	&
	&
	(14--17,17--20) &
	0.668 &
	0 &
	10.45\\
	\hline
	
	4 & 3 Oct 24 & 12:08--12:27 & X9.0 & 0.69
	& 0.70 &
	11--14 &
	1.16 &
	4.88 &
	3.89 &
	(11--14,14--17) &
	0.963 &
	0 &
	6.50\\
	
	& & & &
	& &
	14--17 &
	1.09 &
	4.82 &
	4.12 &
	(11--14,17--20) &
	0.933 &
	0 &
	4.69\\
	
	& & & &
	& &
	17--20 &
	1.09 &
	4.58 &
	3.81 &
	(14--17,17--20) &
	0.971 &
	0 &
	1.08\\
	\hline
	
	5 & 7 Oct 24 & 19:02--19:31 & X2.1 &
	& &
	Noisy data &
	&
	&
	&
	&
	&
	&
	\\
	\hline
	
	6 & 7 Oct 24 & 20:03--21:27 & X1.0 &
	& &
	No QPP &
	&
	&
	&
	&
	&
	&
	\\
	\hline
	
	7 & 9 Oct 24 & 01:25--02:43 & X1.8 & 2.78
	& 0.17 &
	8--11 &
	1.30 &
	0.55 &
	9.74 &
	(8--11,11--14) &
	0.546 &
	0 &
	15.41\\
	
	& & & &
	& &
	11--14 &
	1.84 &
	1.09 &
	11.78 &
	(8--11,14--17) &
	0.452 &
	0 &
	31.89\\
	
	& & & &
	& &
	14--17 &
	1.84 &
	1.62 &
	8.16 &
	(8--11,17--20) &
	0.337 &
	10 &
	28.78\\
	
	& & & &
	& &
	17--20 &
	1.84 &
	4.49 &
	8.16 &
	(11--14,14--17) &
	0.588 &
	0 &
	3.45\\
	
	& & & &
	& &
	&
	&
	&
	&
	(11--14,17--20) &
	0.602 &
	0 &
	11.09\\
	
	& & & &
	& &
	&
	&
	&
	&
	(14--17,17--20) &
	0.667 &
	0 &
	14.94\\
	\hline

	8 & 9 Oct 24 & 15:44--15:53 & X1.4 &
	& &
	No QPP &
	&
	&
	&
	&
	&
	&
	\\
	\hline
	
	9 & 24 Oct 24 & 03:30--04:28 & X3.3 & 1.65
	& 0.50 &
	8--11 &
	1.23 &
	0.63 &
	8.96 &
	(8--11,11--14) &
	0.562 &
	0 &
	5.44\\
	
	& & & &
	& &
	11--14 &
	1.46 &
	0.60 &
	11.99 &
	(8--11,14--17) &
	0.590 &
	0 &
	3.33\\
	
	& & & &
	& &
	14--17 &
	1.23 &
	1.12 &
	5.43 &
	(8--11,17--20) &
	0.501 &
	0 &
	17.53\\
	
	& & & &
	& &
	17--20 &
	1.23 &
	2.05 &
	7.57 &
	(11--14,14--17) &
	0.495 &
	0 &
	-3.83\\
	
	& & & &
	& &
	&
	&
	&
	&
	(11--14,17--20) &
	0.517 &
	0 &
	1.00\\
	
	& & & &
	& &
	&
	&
	&
	&
	(14--17,17--20) &
	0.617 &
	0 &
	13.73\\
	\hline
	
	10 & 26 Oct 24 & 06:32--07:56 & X1.8 &
	& &
	Data N/A &
	&
	&
	&
	&
	&
	&
	\\
	\hline
	
	11 & 31 Oct 24 & 21:12--21:27 & X2.0 & 0.37
	& 0.16 &
	8--11 &
	1.64 &
	2.16 &
	5.29 &
	(8--11,11--14) &
	0.879 &
	0 &
	12.63\\
	
	& & & &
	& &
	11--14 &
	1.74 &
	2.78 &
	4.51 &
	(8--11,14--17) &
	0.850 &
	10 &
	20.92\\
	
	& & & &
	& &
	14--17 &
	1.74 &
	4.03 &
	4.61 &
	(8--11,17--20) &
	0.759 &
	10 &
	36.73\\
	
	& & & &
	& &
	17--20 &
	1.64 &
	8.18 &
	4.38 &
	(11--14,14--17) &
	0.905 &
	0 &
	10.19\\
	
	& & & &
	& &
	&
	&
	&
	&
	(11--14,17--20) &
	0.867 &
	10 &
	25.05\\
	
	& & & &
	& &
	&
	&
	&
	&
	(14--17,17--20) &
	0.922 &
	0 &
	11.80\\
	\hline
	
	12 & 6 Nov 24 & 13:24--13:46 & X2.3 &
	& &
	No QPP &
	&
	&
	&
	&
	&
	&
	\\
	\hline
	
	13 & 8 Dec 24 & 08:50--09:10 & X2.2 &
	& &
	Data N/A &
	&
	&
	&
	&
	&
	&
	\\
	\hline
	
	14 & 29 Dec 24 & 07:08--07:34 & X1.1 & 0.93
	& 0.17 &
	8--11 &
	1.74 &
	2.16 &
	6.82 &
	(8--11,11--14) &
	0.793 &
	0 &
	5.30\\
	
	& & & &
	& &
	11--14 &
	1.74 &
	1.63 &
	5.90 &
	(8--11,14--17) &
	0.759 &
	0 &
	13.41\\
	
	& & & &
	& &
	14--17 &
	1.74 &
	2.36 &
	4.03 &
	(8--11,17--20) &
	0.624 &
	10 &
	57.72\\
	
	& & & &
	& &
	17--20 &
	1.64 &
	3.41 &
	4.99 &
	(11--14,14--17) &
	0.844 &
	0 &
	7.86\\
	
	& & & &
	& &
	&
	&
	&
	&
	(11--14,17--20) &
	0.782 &
	0 &
	28.57\\
	
	& & & &
	& &
	&
	&
	&
	&
	(14--17,17--20) &
	0.853 &
	0 &
	7.95\\
	\hline
	
	15 & 30 Dec 24 & 04:01--04:28 & X1.5 & 0.55
	& 0.17 &
	8--11 &
	1.64 &
	1.12 &
	3.97 &
	(8--11,11--14) &
	0.601 &
	0 &
	15.67\\
	
	& & & &
	& &
	11--14 &
	1.16 &
	1.10 &
	10.50 &
	(8--11,14--17) &
	0.555 &
	0 &
	29.36\\
	
	& & & &
	& &
	14--17 &
	1.23 &
	1.80 &
	5.30 &
	(8--11,17--20) &
	0.440 &
	10 &
	65.96\\
	
	& & & &
	& &
	17--20 &
	1.30 &
	1.87 &
	4.36 &
	(11--14,14--17) &
	0.716 &
	0 &
	6.86\\
	
	& & & &
	& &
	&
	&
	&
	&
	(11--14,17--20) &
	0.647 &
	0 &
	24.46\\
	
	& & & &
	& &
	&
	&
	&
	&
	(14--17,17--20) &
	0.792 &
	0 &
	28.72\\
	\hline
	
	16 & 30 Dec 24 & 04:29--04:34 & X1.1 &
	& &
	No QPP &
	&
	&
	&
	&
	&
	&
	\\
	\hline
	
	17 & 3 Jan 25 & 11:29--11:49 & X1.2 & 0.37
	& 0.52 &
	8--11 &
	1.64 &
	1.20 &
	10.65 &
	(8--11,11--14) &
	0.627 &
	10 &
	17.28\\
	
	& & & &
	& &
	11--14 &
	1.64 &
	1.27 &
	3.09 &
	(8--11,14--17) &
	0.631 &
	10 &
	29.37\\
	
	& & & &
	& &
	14--17 &
	2.92 &
	1.53 &
	9.37 &
	(8--11,17--20) &
	0.222 &
	0 &
	-5.67\\
	
	& & & &
	& &
	17--20 &
	1.64 &
	2.33 &
	7.22 &
	(11--14,14--17) &
	0.615 &
	0 &
	22.51\\
	
	& & & &
	& &
	&
	&
	&
	&
	(11--14,17--20) &
	0.459 &
	30 &
	113.30\\
	
	& & & &
	& &
	&
	&
	&
	&
	(14--17,17--20) &
	0.330 &
	-10 &
	-2.05\\
	\hline
	
	18 & 3 Jan 25 & 22:32--22:51 & X1.1 & 0.35
	& 0.20 &
	8--11 &
	1.95 &
	3.14 &
	4.36 &
	(8--11,11--14) &
	0.869 &
	0 &
	18.40\\
	
	& & & &
	& &
	11--14 &
	1.95 &
	3.91 &
	4.36 &
	(8--11,14--17) &
	0.803 &
	10 &
	50.12\\
	
	& & & &
	& &
	14--17 &
	2.06 &
	4.67 &
	3.80 &
	(8--11,17--20) &
	0.697 &
	20 &
	63.19\\
	
	& & & &
	& &
	17--20 &
	2.06 &
	7.06 &
	4.20 &
	(11--14,14--17) &
	0.823 &
	10 &
	30.77\\
	
	& & & &
	& &
	&
	&
	&
	&
	(11--14,17--20) &
	0.684 &
	10 &
	46.69\\
	
	& & & &
	& &
	&
	&
	&
	&
	(14--17,17--20) &
	0.821 &
	10 &
	12.51\\
	\hline
	
	19 & 4 Jan 25 & 12:34--12:56 & X1.8 & 0.31
	& 0.2 &
	8--11 &
	1.64 &
	1.03 &
	6.82 &
	(8--11,11--14) &
	0.773 &
	0 &
	-2.89\\
	
	& & & &
	& &
	11--14 &
	1.46 &
	1.43 &
	7.42 &
	(8--11,14--17) &
	0.704 &
	0 &
	-22.15\\
	
	& & & &
	& &
	14--17 &
	1.38 &
	1.79 &
	7.14 &
	(8--11,17--20) &
	0.519 &
	0 &
	-20.20\\
	
	& & & &
	& &
	17--20 &
	1.38 &
	2.50 &
	7.02 &
	(11--14,14--17) &
	0.780 &
	0 &
	-17.48\\
	
	& & & &
	& &
	&
	&
	&
	&
	(11--14,17--20) &
	0.634 &
	0 &
	-12.73\\
	
	& & & &
	& &
	&
	&
	&
	&
	(14--17,17--20) &
	0.749 &
	0 &
	-1.49\\
	\hline
	
	20 & 23 Feb 25 & 19:22--19:34 & X2.0 &
	& &
	No QPP &
	&
	&
	&
	&
	&
	&
	\\
	\hline
	
	21 & 28 Mar 25 & 15:03--15:42 & X1.1 & 0.74
	& 0.35 &
	8--11 &
	1.38 &
	1.06 &
	5.57 &
	(8--11,11--14) &
	0.744 &
	0 &
	12.45\\
	
	& & & &
	& &
	11--14 &
	1.38 &
	3.06 &
	9.32 &
	(8--11,14--17) &
	0.637 &
	0 &
	46.82\\
	
	& & & &
	& &
	14--17 &
	1.30 &
	6.03 &
	10.00 &
	(8--11,17--20) &
	0.504 &
	0 &
	71.84\\
	
	& & & &
	& &
	17--20 &
	1.30 &
	8.81 &
	9.23 &
	(11--14,14--17) &
	0.797 &
	0 &
	20.99\\
	
	& & & &
	& &
	&
	&
	&
	&
	(11--14,17--20) &
	0.713 &
	0 &
	38.01\\
	
	& & & &
	& &
	&
	&
	&
	&
	(14--17,17--20) &
	0.908 &
	0 &
	4.23\\
	\hline

	22 & 13 May 25 & 15:25--15:44 & X1.2 & 0.62
	& 0.64 &
	8--11 &
	2.19 &
	2.88 &
	4.19 &
	(8--11,11--14) &
	0.866 &
	0 &
	2.02\\
	
	& & & &
	& &
	11--14 &
	2.19 &
	3.63 &
	4.12 &
	(8--11,14--17) &
	0.671 &
	10 &
	42.68\\
	
	& & & &
	& &
	14--17 &
	2.32 &
	3.50 &
	4.03 &
	(8--11,17--20) &
	0.679 &
	40 &
	108.82\\
	
	& & & &
	& &
	17--20 &
	2.32 &
	6.51 &
	3.60 &
	(11--14,14--17) &
	0.809 &
	10 &
	52.61\\
	
	& & & &
	& &
	&
	&
	&
	&
	(11--14,17--20) &
	0.789 &
	50 &
	127.29\\
	
	& & & &
	& &
	&
	&
	&
	&
	(14--17,17--20) &
	0.646 &
	40 &
	31.10\\
	\hline
	
	23 & 14 May 25 & 08:04--08:31 & X2.7 &
	& &
	No QPP &
	&
	&
	&
	&
	&
	&
	\\
	\hline
	
	24 & 25 May 25 & 01:46--01:57 & X1.1 & 0.39
	& 0.18 &
	11--14 &
	1.74 &
	3.08 &
	4.61 &
	(11--14,14--17) &
	0.904 &
	0 &
	8.12\\
	
	& & & &
	& &
	14--17 &
	1.74 &
	3.16 &
	6.24 &
	(11--14,17--20) &
	0.676 &
	10 &
	17.20\\
	
	& & & &
	& &
	17--20 &
	1.74 &
	3.46 &
	5.28 &
	(14--17,17--20) &
	0.612 &
	0 &
	15.85\\
	\hline
	
	25 & 17 Jun 25 & 21:38--21:54 & X1.2 & 0.93
	& 0.39 &
	8--11 &
	1.38 &
	3.53 &
	8.85 &
	(8--11,11--14) &
	0.850 &
	0 &
	5.68\\
	
	& & & &
	& &
	11--14 &
	1.38 &
	5.30 &
	7.26 &
	(8--11,14--17) &
	0.708 &
	10 &
	7.09\\
	
	& & & &
	& &
	14--17 &
	1.55 &
	7.99 &
	4.74 &
	(8--11,17--20) &
	0.452 &
	10 &
	48.80\\
	
	& & & &
	& &
	17--20 &
	1.64 &
	8.58 &
	4.78 &
	(11--14,14--17) &
	0.837 &
	10 &
	22.70\\
	
	& & & &
	& &
	&
	&
	--- &
	&
	(11--14,17--20) &
	0.679 &
	10 &
	31.57\\
	
	& & & &
	& &
	&
	&
	&
	&
	(14--17,17--20) &
	0.828 &
	0 &
	41.33\\
	\hline
	
	26 & 19 Jun 25 & 23:37--23:54 & X1.9 &
	& &
	Data N/A &
	&
	&
	&
	&
	&
	&
	\\
	\hline
	
	27 & 4 Nov 25 & 17:15--17:51 & X1.8 & 1.10
	& 0.27 &
	8--11 &
	1.23 &
	0.50 &
	17.66 &
	(8--11,11--14) &
	0.529 &
	0 &
	-15.52\\
	
	& & & &
	& &
	11--14 &
	1.74 &
	0.82 &
	9.60 &
	(8--11,14--17) &
	0.310 &
	0 &
	2.64\\
	
	& & & &
	& &
	14--17 &
	1.23 &
	0.93 &
	18.85 &
	(8--11,17--20) &
	0.471 &
	0 &
	8.06\\
	
	& & & &
	& &
	17--20 &
	1.38 &
	1.96 &
	13.43 &
	(11--14,14--17) &
	0.574 &
	0 &
	6.39\\
	
	& & & &
	& &
	&
	&
	&
	&
	(11--14,17--20) &
	0.367 &
	0 &
	25.95\\
	
	& & & &
	& &
	&
	&
	&
	&
	(14--17,17--20) &
	0.464 &
	0 &
	19.71\\
	\hline
	
	28 & 4 Nov 25 & 21:45--22:11 & X1.1 & 1.75
	& 0.89 &
	8--11 &
	1.74 &
	1.01 &
	10.51 &
	(8--11,11--14) &
	0.746 &
	0 &
	13.53\\
	
	& & & &
	& &
	11--14 &
	1.74 &
	1.60 &
	8.26 &
	(8--11,14--17) &
	0.610 &
	0 &
	45.30\\
	
	& & & &
	& &
	14--17 &
	1.46 &
	2.35 &
	8.68 &
	(8--11,17--20) &
	0.633 &
	10 &
	42.62\\
	
	& & & &
	& &
	17--20 &
	1.46 &
	3.05 &
	9.44 &
	(11--14,14--17) &
	0.604 &
	0 &
	29.83\\
	
	& & & &
	& &
	&
	&
	&
	&
	(11--14,17--20) &
	0.648 &
	10 &
	32.59\\
	
	& & & &
	& &
	&
	&
	&
	&
	(14--17,17--20) &
	0.606 &
	10 &
	9.91\\
	\hline

	29 & 9 Nov 25 & 07:01--07:55 & X1.7 & 1.31
	& 0.49 &
	8--11 &
	1.30 &
	0.90 &
	8.23 &
	(8--11,11--14) &
	0.632 &
	0 &
	-4.73\\
	
	& & & &
	& &
	11--14 &
	1.38 &
	1.20 &
	8.27 &
	(8--11,14--17) &
	0.686 &
	0 &
	-4.73\\
	
	& & & &
	& &
	14--17 &
	1.38 &
	1.60 &
	7.14 &
	(8--11,17--20) &
	0.615 &
	0 &
	1.11\\
	
	& & & &
	& &
	17--20 &
	1.38 &
	2.58 &
	10.49 &
	(11--14,14--17) &
	0.636 &
	0 &
	0.00\\
	
	& & & &
	& &
	&
	&
	&
	&
	(11--14,17--20) &
	0.480 &
	0 &
	11.27\\
	
	& & & &
	& &
	&
	&
	&
	&
	(14--17,17--20) &
	0.664 &
	0 &
	11.27\\
	\hline
	
	30 & 10 Nov 25 & 08:55--10:19 & X1.2 & 1.56
	& 0.39 &
	8--11 &
	2.92 &
	0.97 &
	8.99 &
	(8--11,11--14) &
	0.707 &
	0 &
	-0.08\\
	
	& & & &
	& &
	11--14 &
	2.92 &
	2.28 &
	7.48 &
	(8--11,14--17) &
	0.733 &
	10 &
	19.23\\
	
	& & & &
	& &
	14--17 &
	2.92 &
	2.16 &
	7.37 &
	(8--11,17--20) &
	0.628 &
	10 &
	9.37\\
	
	& & & &
	& &
	17--20 &
	2.92 &
	3.22 &
	8.17 &
	(11--14,14--17) &
	0.793 &
	0 &
	3.78\\
	
	& & & &
	& &
	&
	&
	&
	&
	(11--14,17--20) &
	0.784 &
	0 &
	7.17\\
	
	& & & &
	& &
	&
	&
	&
	&
	(14--17,17--20) &
	0.804 &
	0 &
	-3.84\\
	\hline
	
	31 & 11 Nov 25 & 09:49--10:17 & X5.1 &
	& &
	No QPP &
	&
	&
	&
	&
	&
	&
	\\
	\hline
	
	32 & 14 Nov 25 & 07:44--08:40 & X4.0 & 1.65
	& 0.43 &
	8--11 &
	1.64 &
	0.76 &
	8.45 &
	(8--11,11--14) &
	0.676 &
	0 &
	-14.91\\
	
	& & & &
	& &
	11--14 &
	1.74 &
	1.54 &
	8.34 &
	(8--11,14--17) &
	0.709 &
	0 &
	16.26\\
	
	& & & &
	& &
	14--17 &
	1.64 &
	1.44 &
	8.04 &
	(8--11,17--20) &
	0.655 &
	0 &
	17.62\\
	
	& & & &
	& &
	17--20 &
	1.55 &
	1.77 &
	8.41 &
	(11--14,14--17) &
	0.800 &
	0 &
	28.07\\
	
	& & & &
	& &
	&
	&
	&
	&
	(11--14,17--20) &
	0.804 &
	10 &
	27.70\\
	
	& & & &
	& &
	&
	&
	&
	&
	(14--17,17--20) &
	0.827 &
	0 &
	4.41\\
	\hline
	
	33 & 1 Dec 25 & 02:27--03:05 & X1.9 & 0.62
	& 0.64 &
	8--11 &
	2.45 &
	0.68 &
	10.04 &
	(8--11,11--14) &
	0.560 &
	0 &
	10.80\\
	
	& & & &
	& &
	11--14 &
	1.38 &
	1.28 &
	7.02 &
	(8--11,14--17) &
	0.487 &
	10 &
	40.30\\
	
	& & & &
	& &
	14--17 &
	1.38 &
	1.22 &
	11.13 &
	(8--11,17--20) &
	0.326 &
	0 &
	27.80\\
	
	& & & &
	& &
	17--20 &
	2.06 &
	1.68 &
	7.03 &
	(11--14,14--17) &
	0.561 &
	0 &
	-30.45\\
	
	& & & &
	& &
	&
	&
	&
	&
	(11--14,17--20) &
	0.378 &
	0 &
	-36.87\\
	
	& & & &
	& &
	&
	&
	&
	&
	(14--17,17--20) &
	0.509 &
	0 &
	-14.53\\
	\hline
	
	34 & 8 Dec 25 & 04:49--05:04 & X1.1 &
	& &
	No QPP &
	&
	&
	&
	&
	&
	&
	\\
	\hline
	
	35 & 18 Jan 26 & 11:42--12:18 & X2.3 & 0.74
	& 0.15 &
	8--11 &
	2.06 &
	0.54 &
	5.73 &
	(8--11,11--14) &
	0.541 &
	0 &
	9.13\\
	
	& & & &
	& &
	11--14 &
	2.06 &
	1.20 &
	9.45 &
	(8--11,14--17) &
	0.499 &
	10 &
	28.79\\
	
	& & & &
	& &
	14--17 &
	1.09 &
	1.51 &
	15.70 &
	(8--11,17--20) &
	0.435 &
	10 &
	65.75\\
	
	& & & &
	& &
	17--20 &
	1.09 &
	2.98 &
	11.13 &
	(11--14,14--17) &
	0.443 &
	10 &
	25.28\\
	
	& & & &
	& &
	&
	&
	&
	&
	(11--14,17--20) &
	0.347 &
	10 &
	60.85\\
	
	& & & &
	& &
	&
	&
	&
	&
	(14--17,17--20) &
	0.537 &
	0 &
	6.30\\
	\hline
	
	36 & 1 Feb 26 & 12:25--12:37 & X1.0 &
	& &
	Data N/A &
	&
	&
	&
	&
	&
	&
	\\
	\hline
	
	37 & 1 Feb 26 & 23:44--00:04 & X8.1 &
	& &
	No QPP &
	&
	&
	&
	&
	&
	&
	\\
	\hline
	
	38 & 2 Feb 26 & 00:08--00:32 & X1.5 & --
	& -- &
	8--11 &
	1.84 &
	2.13 &
	4.53 &
	(8--11,11--14) &
	0.894 &
	0 &
	14.51\\
	
	& & & &
	& &
	11--14 &
	1.64 &
	3.11 &
	4.13 &
	(8--11,14--17) &
	0.840 &
	0 &
	21.33\\
	
	& & & &
	& &
	14--17 &
	2.45 &
	4.46 &
	3.33 &
	(8--11,17--20) &
	0.836 &
	0 &
	24.58\\
	
	& & & &
	& &
	17--20 &
	1.46 &
	5.55 &
	3.27 &
	(11--14,14--17) &
	0.934 &
	0 &
	1.40\\
	
	& & & &
	& &
	&
	&
	&
	&
	(11--14,17--20) &
	0.933 &
	0 &
	0.87\\
	
	& & & &
	& &
	&
	&
	&
	&
	(14--17,17--20) &
	0.978 &
	0 &
	0.88\\
	\hline
	
	39 & 2 Feb 26 & 08:14--08:43 & X2.7 & --
	& -- &
	8--11 &
	1.38 &
	0.45 &
	7.89 &
	(8--11,11--14) &
	0.493 &
	0 &
	9.99\\
	
	& & & &
	& &
	11--14 &
	1.46 &
	0.85 &
	6.06 &
	(8--11,14--17) &
	0.420 &
	0 &
	44.51\\
	
	& & & &
	& &
	14--17 &
	1.84 &
	0.94 &
	11.42 &
	(8--11,17--20) &
	0.288 &
	10 &
	46.95\\
	
	& & & &
	& &
	17--20 &
	1.09 &
	1.65 &
	7.07 &
	(11--14,14--17) &
	0.459 &
	0 &
	-15.33\\
	
	& & & &
	& &
	&
	&
	&
	&
	(11--14,17--20) &
	0.386 &
	0 &
	2.29\\
	
	& & & &
	& &
	&
	&
	&
	&
	(14--17,17--20) &
	0.574 &
	0 &
	4.84\\
	\hline
	
	40 & 3 Feb 26 & 13:58--14:18 & X1.5 &
	& &
	Data N/A &
	&
	&
	&
	&
	&
	&
	\\
	\hline
	
	41 & 4 Feb 26 & 12:02--12:18 & X4.2 &
	& &
	Data N/A &
	&
	&
	&
	&
	&
	&
	\\
	\hline
	
\end{longtable}
	

\end{document}